\documentclass[rnote]{aa} 

\usepackage{graphicx}
\usepackage{txfonts}
\usepackage{natbib}
\bibpunct{(}{)}{;}{a}{}{,}
\usepackage{color}

\begin{document}

   \title{Doppler-beaming in the Kepler light curve of LHS~6343~A}

\author{E.~Herrero\inst{\ref{inst1}} \and A.~F.~Lanza\inst{\ref{inst2}} \and I.~Ribas\inst{\ref{inst1}} \and 
C.~Jordi\inst{\ref{inst3}} \and A.~Collier~Cameron\inst{\ref{inst4}} \and J.~C.~Morales\inst{\ref{inst5}}
} 
\institute{Institut de Ci\`{e}ncies de l'Espai (CSIC-IEEC), Campus UAB, 
Facultat de Ci\`{e}ncies, Torre C5 parell, 2a pl, 08193 Bellaterra, 
Spain, \email{eherrero@ice.cat, iribas@ice.cat, morales@ice.cat}\label{inst1}
\and
INAF - Osservatorio Astrofisico di Catania, via S. Sofia, 78, 95123 Catania, Italy,
\email{nuccio.lanza@oact.inaf.it}\label{inst2}
\and
Dept. d'Astronomia i Meteorologia, Institut de Ci\`{e}ncies del Cosmos (ICC),
Universitat de Barcelona (IEEC-UB), Mart\'{i} Franqu\`{e}s 1, E08028 Barcelona, Spain, 
\email{carme.jordi@ub.edu}\label{inst3}
\and
SUPA, School of Physics and Astronomy, University of St. Andrews, North Haugh, Fife KY16 9SS,
\email{acc4@st-andrews.ac.uk}\label{inst4}
\and
LESIA-Observatoire de Paris, CNRS, UPMC Univ. Paris 06, Univ. Paris-Diderot, 5 Pl. Jules Janssen, 92195 Meudon CEDEX, France,
\email{Juan-Carlos.Morales@obspm.fr}\label{inst5}
}

\date{Received date /
Accepted date}
 
  \abstract
   {Kepler observations revealed a brown dwarf eclipsing the M-type star LHS~6343~A with a period of 12.71~days. In addition, an out-of-eclipse light modulation with the same period and a relative semi-amplitude of $\sim2 \times 10^{-4}$ was observed showing an almost constant phase lag to the eclipses produced by the brown dwarf.  In a previous work, we concluded that this was due to the light modulation induced by photospheric active regions in LHS~6343~A.}
   {In the present work, we prove that most of the out-of-eclipse light modulation  is caused by the Doppler-beaming induced by the orbital motion of the primary star. }
   {We introduce a model of the Doppler-beaming for an eccentric orbit and also considered the ellipsoidal effect.  The data were fitted using a Bayesian approach implemented through a Monte Carlo Markov chain method. Model residuals were analysed by searching for periodicities using a Lomb-Scargle periodogram.}
   {For the first seven quarters of Kepler observations and the orbit previously derived from the radial velocity measurements, we show that the light modulation of the system outside eclipses is dominated by the Doppler-beaming effect. 
   A period search performed on the residuals shows  a  significant periodicity of $ 42.5 \pm 3.2$~days with a false-alarm probability of $5 \times 10^{-4}$, probably associated with the rotational modulation of the primary component.}
   {}

\maketitle

\section{Introduction}

In addition to the detection of Earth-like exoplanets, the highly accurate photometry provided by the Kepler mission has allowed the community to discover a number of eclipsing binaries and study stellar variability at very low amplitudes. Several detections of flux modulations in binary stars have been associated with relativistic beaming caused by the radial motion of their components \citep{Kerkwijk10,Bloemen11}. The effect is proportional to the orbital velocity of the component stars  and allows one to estimate their radial velocity amplitudes  in selected compact binaries. This photometric method to measure radial velocities was first introduced by \cite{Shakura87} and applied by \cite{Maxted00}. In the context of a possible application to CoRoT and Kepler light curves, it was first discussed by \cite{Loeb03} and \cite{Zucker07}.

In this paper, we present an interpretation of the out-of-eclipse light modulation in the Kepler photometry of LHS~6343 (KID 010002261) in terms of a  Doppler-beaming effect. This eclipsing binary consists of an M4V star (component A) and a brown dwarf (62.7 M$_{\rm Jup}$, component C) and was discovered by \cite{2011ApJ...730...79J} as part of the system LHS~6343~AB, a visual binary consisting of two M-dwarf stars with a projected separation of 0\farcs55. \cite{Herrero13} analysed a more extended  Kepler time series,
which  revealed a modulation in the flux of LHS~6343~A, synchronized with the brown dwarf orbital motion with a minimum preceding the subcompanion point by $\sim100^{\circ}$. These oscillations were assumed to be caused by persistent groups of starspots. A maximum-entropy spot-modelling technique was applied to extract the primary star rotation period, the typical lifetime of the
spot, and some evidence of a possible magnetic interaction to account for the close synchronicity and almost constant phase lag between the modulation and the eclipses.

The Doppler-beaming modelling that we present in this paper shows that the main modulation signal can instead be explained by this effect, and that the radial velocity amplitude as derived from the light curve is compatible with the spectroscopically determined value \citep{2011ApJ...730...79J}. Doppler-beaming has been previously detected in Kepler light curves of KOI-74 and KOI-81 by \cite{Kerkwijk10}, KPD 1946+4340 by \cite{Bloemen11}, KIC 10657664 by \cite{Carter11} and KOI 1224 by \cite{Breton12}.  A proper modelling of the effects observed in the light curves of these objects is important because it may give us the opportunity to derive radial velocities from a number of binaries observed by Kepler and to remove the Doppler-beaming modulation to investigate other causes of light-curve variation.

\section{Photometry}

LHS 6343 (KIC 010002261) was observed by Kepler during its entire mission lifetime. In this work, we re-analyse the same time series as in \cite{Herrero13}, consisting of the first seven quarters of observations (Q0 to Q6). The time series consists of a total of 22976 data points with  $\sim$30 minute cadence, a mean relative precision of $7 \times 10^{-5}$, and spans  $\sim 510$ days ranging from May 2009 to September 2010. The two M-type components A and B of the visual binary, separated by 0\farcs55 \citep{2011ApJ...730...79J}, are contained inside a single pixel of the Kepler images (the pixel side being 3\farcs98), and hence any photometric mask selected for the A component contains contamination from component B. 

Co-trending basis vectors are applied to the raw data using the PyKE pipeline reduction software \footnote{http://keplergo.arc.nasa.gov/PyKE.shtml} to correct for systematic trends, which are mainly  related to the pointing jitter of the satellite, detector instabilities, and environment variations \citep{Murphy12}. These are optimised tasks to reduce Kepler Simple Aperture Photometry (hereafter SAP) data\footnote{The SAP light curve is a pixel summation time-series of the entire calibrated flux that falls within the optimal aperture.} because they account for the position of the specific target on the detector plane to correct for systematics.  From two to four vectors are used for each quarter to remove the main trends from the raw data. A low-order ($\leq 4 $) polynomial filtering is then applied to the resulting data for each quarter because some residual trends still remain, which are followed by discontinuities between quarters. These are due to the change of the target position on the focal plane following  each re-orientation of the spacecraft at the end of each quarter.  As a consequence of this data reduction process, the general trends disappear, and the use of low-order polynomials ensures that the frequency and amplitude of any variability with a time scale $\la 50$ days is preserved. Several gaps in the data prevent us from using other detrending methods such as Fourier filtering \citep[cf., ][]{Herrero13}.

\begin{figure*}[t]
\centerline{
\includegraphics[width=8.5cm,angle=-90]{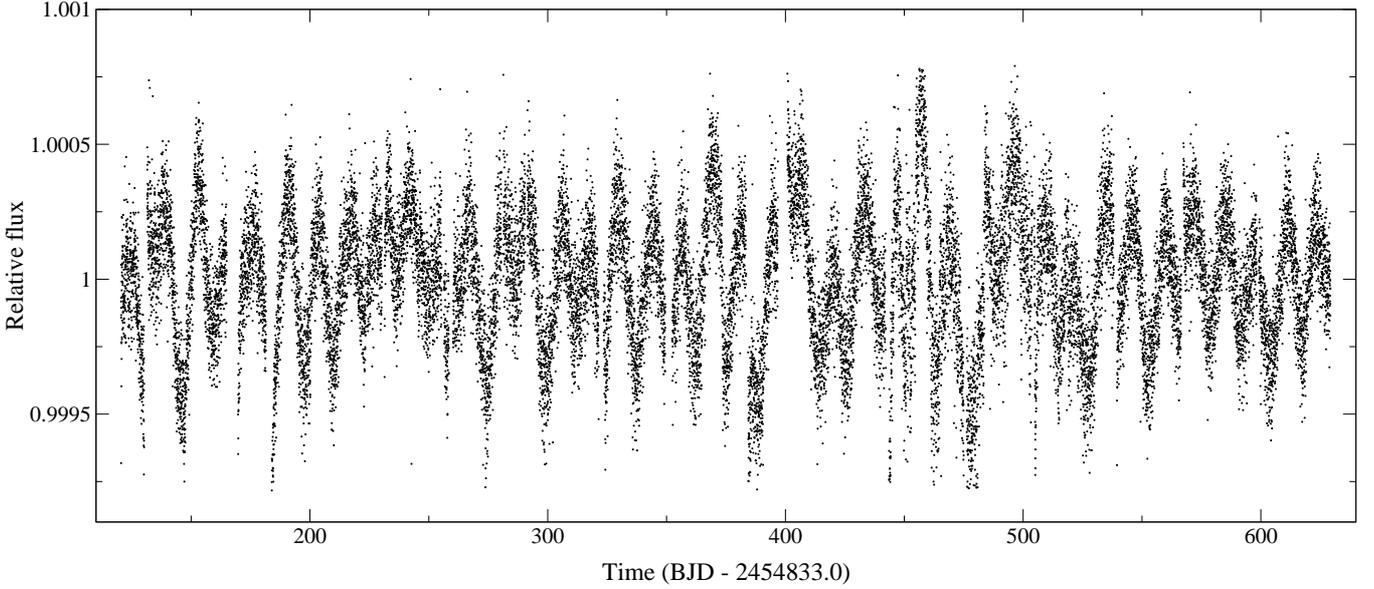}}  
\caption{Out-of-eclipse Kepler light curve of LHS 6343 A, covering the first seven quarters, after detrending and flux dilution correction as adopted for the subsequent analysis.}
\label{lcraw}
\end{figure*}

The contamination from component B was corrected by subtracting its flux contribution before modelling the light curve. \cite{2011ApJ...730...79J} used independent Johnson-V photometry of the A and B components together with stellar models  to estimate the magnitude difference in the Kepler passband, obtaining $\Delta K_{P}=0.74\pm0.10$. This is equivalent to a flux ratio of $1.97\pm0.19$, which was applied to correct for the flux dilution produced by component B. Finally, eclipses were removed from the data set considering the ephemeris and the system parameters in  \cite{Herrero13}. A complete analysis of the photometry of the brown dwarf eclipses can be found in \cite{2011ApJ...730...79J} and \cite{Herrero13}. The de-trended out-of-eclipse light curve is shown in Fig.\ref{lcraw}, while raw SAP data have been  presented in Fig. 1 of \cite{Herrero13}.

\section{Models of the Doppler-beaming and ellipticity effect}

A first-order approximation in $v_{\rm R}/c$ for the flux variation  at frequency $\nu$ due to Doppler-beaming is \citep[cf.][]{Rybicki79,Zucker07}
\begin{equation}
\left.\frac{\Delta F}{F}\right|_{\rm DB} = - (3-\alpha) \frac{v_{\rm R}(t)}{c},
\label{db_effect}
\end{equation}
where $v_{\rm R}(t)$ is the radial velocity of the star at time $t$, $c$ the speed of light, and the spectral index $\alpha \equiv d\ln F_{\nu}/d \ln \nu$ depends on the spectrum of the star $F_{\nu}$. Doppler-beaming produces an increase of the bolometric flux for a source that is approaching the observer, that is, when $v_{\rm R} < 0$.
In the case of LHS 6343 A, the Doppler shift of  the radiation towards the blue when the star is approaching the observer causes the flux in the Kepler passband to increase because we observe photons with a longer wavelength in the rest frame of the source, which corresponds to a higher flux, given the low effective temperature of the star ($T_{\rm eff} \sim 3000$~K). In other words, $\alpha < 0$ for a star as cool as LHS~6343~A. 

We computed a mean spectral index by integrating BT-Settl model spectra \citep{Allard11} over the photon-weighted Kepler passband,
\begin{equation}
\langle \alpha \rangle = \frac{\int h_{\nu} \nu F_{\nu} \alpha d\nu}{\int h_{\nu} \nu F_{\nu} d\nu},
\label{sp_index}
\end{equation}
where $h_{\nu}$ is the response function of the Kepler passband. For the stellar model $F_{\nu}$ we assumed solar metallicity, $T_{\rm eff}=3130$~K, $\log g=4.851$ (cm s$^{-2}$) and an $\alpha$-element enhancement $[\alpha/\rm H]=0$. The resulting mean spectral index to be used in Eq. (\ref{db_effect}) is $\langle \alpha \rangle = -3.14 \pm 0.08$. The uncertainty comes from the dependence of the spectral index on the model spectrum and the uncertainties of the respective parameters and is evaluated by calculating the integral in Eq.~(\ref{sp_index}) by varying the temperature in the range $T_{\rm eff}=3130\pm20$~K and the surface gravity in $\log g=4.851\pm0.008$ (cm s$^{-2}$) \citep{2011ApJ...730...79J}. If we compute the spectral index considering the black-body approximation \citep{Zucker07}, the result is $\langle \alpha_{\rm BB} \rangle \simeq -4.22$  for the Kepler passband. The difference is due to the many absorption features in the spectrum of this type of stars that fall within the Kepler passband.

Assuming a reference frame with the origin at the barycentre of the binary system and the $z-$ axis pointing away from the observer, we can express the radial velocity of the primary component as a trigonometric series in  the mean anomaly $M$ by applying the elliptic expansions reported in \citet{MurrayDermott99}:
\begin{eqnarray}
v_{\rm R} & = & A\left[\left( 1- \frac{9e^{2}}{8}\right) \cos M + \left(e - \frac{4e^{3}}{3} \right) \cos 2M  \right. \nonumber \\
   & & + \left. \frac{9e^{2}}{8} \cos 3M + \frac{4e^{3}}{3} \cos 4M \right] \nonumber \\
   & + & B \left[ \left( 1- \frac{7e^{2}}{8}\right) \sin M + \left(e - \frac{7e^{3}}{6} \right) \sin 2M  \right. \nonumber \\
 & & \left. + \frac{9e^{2}}{8} \sin 3M + \frac{4e^{3}}{3} \sin 4M \right] + O (e^{4}),
 \label{vr_eq}
\end{eqnarray}
where $A \equiv K \cos \omega$ and $B\equiv -K \sin \omega$, with $K$ the radial velocity semi-amplitude, $\omega$ the argument of periastron, and $e$ the eccentricity of the orbit of the primary component  \citep[see, e.g., ][]{WrightHoward09}. At the epoch of  mid-eclipse of the primary by  the brown dwarf, the true anomaly is \citep[cf., e.g., ][]{Winn11}: $f_{\rm e} = \frac{\pi}{2}-\omega$. From the true anomaly at mid-eclipse, we find the eccentric anomaly and the mean anomaly:
\begin{equation}
\tan \frac{E_{\rm e}}{2} = \sqrt{\frac{1-e}{1+e}} \tan \frac{f_{\rm e}}{2}, 
\end{equation}
and 
\begin{equation}
M_{\rm e} = E_{\rm e} - e \sin E_{\rm e}.
\end{equation}
If we measure the time since the mid-eclipse epoch $T_{0}$, the mean anomaly appearing in Eq. (\ref{vr_eq}) is
\begin{equation}
M = n (t-T_{0}) + M_{\rm e}, 
\end{equation}
because $M$ is zero at the epoch of periastron. 

In addition to the Doppler-beaming effect, the ellipsoidal effect can be important in the case of LHS~6343~A, while the reflection effect is negligible because of a relative separation of $\sim 45.3$ stellar radii in the system and the low luminosity of the C secondary component. \cite{Morris85} provided formulae to evaluate the effect. In our case, only the coefficient proportional to $\cos 2 \phi$, where $\phi = M - M_{\rm e}$ is the orbital angular phase, is relevant because the 
other terms are at least one order of magnitude smaller due to the large relative separation. In terms of the mean anomaly, the relative flux modulation due to the ellipsoidal effect is
\begin{eqnarray}
\left.\frac{\Delta F}{F}\right|_{\rm E} & = & C_{1}(2) \cos (2M -2M_{\rm e}) = \nonumber \\
  &  = & [C_{1}(2)\sin(2M_{\rm e})]\sin 2M + [C_{1}(2)\cos(2M_{\rm e})] \cos  2M, \nonumber \\
 & &
 \label{elip1}
\end{eqnarray}
where
\begin{equation}
C_{1}(2) = - Z_{1}(2) \left( \frac{m}{M_{*}} \right) \left( \frac{R}{a} \right)^{3} \sin^{2} i,
\label{elip2}
\end{equation}
$m$ is the mass of the brown dwarf secondary, $M_{*}$ the mass of the distorted  primary star, $R$ its radius, and $i$ the inclination of the orbital plane,  which are fixed to those derived by \cite{2011ApJ...730...79J}. Finally, $Z_{1}(2)$ is a coefficient given by
\begin{equation}
Z_{1}(2) \simeq \frac{45 + 3u}{20(3-u)}  \left( \tau_{\rm g}  +1 \right),
\label{elip3}
\end{equation}
where $u=1.2$ is the linear limb-darkening coefficient in the Kepler passband, $\tau_{\rm g} \sim 0.32$  the gravity-darkening coefficient estimated for the primary LHS~6343~A, and we neglected the effects related to the precession constant and the (small) eccentricity of the orbit \citep[cf., ][]{Morris85}. Note that  at mid-eclipse, $M=M_{\rm e}$, and the ellipsoidal variation is at a minimum, while for 
$M=M_{\rm e} \pm \pi/2$, that is, in quadrature, it reaches a maximum. 

In conclusion, the total relative light variation due to both Doppler-beaming and ellipsoidal effect is
\begin{equation}
\frac{\Delta F}{F} = \left.\frac{\Delta F}{F}\right|_{\rm DB} + \left.\frac{\Delta F}{F}\right|_{\rm E}.  
\end{equation}

\section{Results}

To fit the proposed model to the data, we applied a Monte Carlo Markov chain  (MCMC) approach that allowed us to find, in addition to the best-fit solution, the posterior  distribution of the parameters that provides us with their uncertainties and correlations. We followed the method outlined in Appendix~A of \citet{Sajinaetal06} \citep[see also][]{Pressetal02,Ford06}. If $\vec a \equiv \{ e, \omega, K\}$ is the vector of the parameter values, and $\vec d$ the vector of the data points, according to the Bayes theorem we have 
\begin{equation}
p({\vec a} | {\vec d} ) \propto p ( {\vec d} | {\vec a}) p ({\vec a}), 
\label{post_prob}
\end{equation}
where $p({\vec a} | {\vec d})$ is the a posteriori probability distribution of the parameters, $p({\vec d} | {\vec a})$  the likelihood of the data for the given model, and $p({\vec a})$  the prior. In our case, the parameters  have been derived by \cite{2011ApJ...730...79J} by fitting the radial velocity and  transit light-curves. Therefore,  we can use their values and uncertainties to define the prior  as
\begin{eqnarray}
p(\vec a ) & = &\exp\left\{ -\left [\frac{(e-0.056)}{0.032} \right]^{2} - \left[\frac{(\omega +23)}{56} \right]^{2}  \right. \nonumber \\
 & & \left. - \left[ \frac{(K-9.6)}{0.3} \right]^{2} \right\},
 \label{prior}
\end{eqnarray}
where $\omega$ is measured in degrees and $K$ in km~s$^{-1}$. The likelihood of the data for given model parameters is 
\begin{equation}
p({\vec d} | {\vec a}) \propto  \exp (-\chi^{2}_{\rm r}), 
\label{p_da}
\end{equation}
where $\chi^{2}_{\rm r}$ is the reduced chi-square of the fit to the data obtained with our model.  The standard deviation of the data used to compute $\chi^{2}_{\rm r}$ is the mean of the standard deviations evaluated in 40 equal bins of the mean anomaly and is $\sigma_{\rm m} = 2.057 \times 10^{-4}$ in relative flux units. Note that in addition to estimating the standard deviation of the data, we always fitted the unbinned time-series shown in 
Fig.~\ref{lcraw}. Substituting Eqs.~(\ref{p_da}) and~(\ref{prior}) into Eq.~(\ref{post_prob}), we obtain the posterior probability distribution of the parameters. We sampled from this distribution by means of the Metropolis-Hasting algorithm \citep[cf., e.g., ][]{Pressetal02}, thus avoiding the complicated problem of normalizing $p({\vec a} | {\vec d})$ over a multi-dimensional parameter space. A Monte Carlo Markov chain is built by performing a conditioned random walk within the parameter space. Specifically, starting from a given point ${\vec a}_{i}$, a proposal is made to move to a successive point ${\vec a}_{i+1}$ whose coordinates are found by incrementing  those of the initial point by random deviates taken from a multi-dimensional Gaussian distribution with standard deviations $\sigma_{j}$, where $j=1,2$ or 3 indicate the parameter. With this choice for the proposed increments of the parameters, the step is accepted if $p({\vec a}_{i+1} | {\vec d})/p({\vec a}_{i} | {\vec d}) > u$, where $u$ is a random number between 0 and 1 drawn from a uniform distribution, otherwise we return to the previous point, that is, ${\vec a}_{i+1} = {\vec a}_{i}$. 

We computed a chain of 200 000 points adjusting $\sigma_{j}$  to have an average acceptance probability of 23 percent  that guarantees a proper sampling and minimises the internal correlation of the chain itself.  The mixing and convergence of the chain to the posterior parameter distribution were tested with the method of Gelman and Rubin as implemented by \citet{Verdeetal03}. First we discarded the first 25 000 points that correspond to the initial phase during which the Metropolis-Hasting algorithm converges on the stationary final distribution (the so-called burn-in phase), then we cut the remaining chain  into four subchains that were used to compute the test parameter $R$. It must be lower than 1.1 when the chain has converged on the distribution to be sampled. In our case we found $(R -1) \leq 4.4 \times 10^{-4}$ for all the three parameters, which indicates convergence and good sampling of the parameter space. 

The best-fit model  corresponding to the minimum $\chi^{2}_{\rm r} = 1.0033$ has the parameters $e=0.0448$, $\omega=-90\fdg704$, and $K=9.583$ km~s$^{-1}$.
For comparison, the reduced chi-square corresponding to the best-fit parameters of  \cite{2011ApJ...730...79J} is $1.0098$. 

\begin{figure}[]
\centerline{
\includegraphics[height=9cm,angle=90]{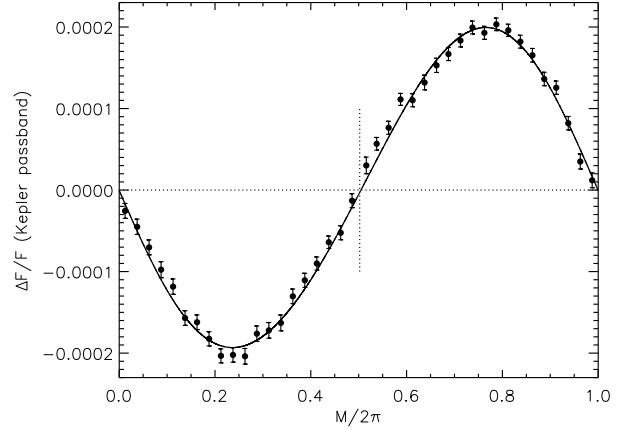}} 
\caption{Relative flux variation of LHS 6343 A  vs. the mean anomaly $M$ of the orbit, binned in 40 equal intervals. The  Doppler-beaming plus ellipsoidal effect model for our best-fitting  parameters  is plotted with a solid line (cf. the text). The value of $M/2\pi$ corresponding to mid-eclipse is marked with a dotted vertical segment, while a horizontal dotted line is plotted to indicate the zero-flux level. }
\label{eccentricfit}
\end{figure}

\begin{figure}[]
\centerline{
\includegraphics[height=9cm,width=8cm,angle=0]{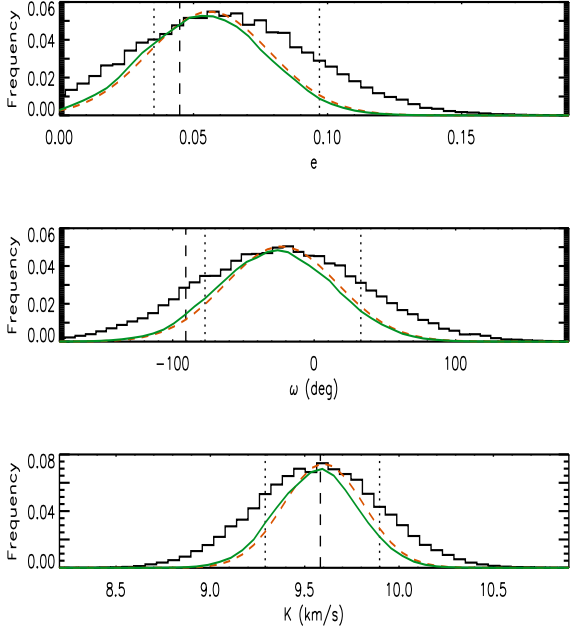}}

\caption{Top panel: The histogram shows the a posteriori distribution of 
eccentricity $e$ as obtained with our MCMC approach. The vertical dashed 
line indicates the value corresponding to the best fit  plotted in 
Fig.~\ref{eccentricfit}, while the two vertical dotted lines enclose an 
interval between 15.9 and 84.1 percents of the distribution.
The solid green line is the mean likelihood as computed by means of 
Eq.~(A4) of \citet{Sajinaetal06}, while the orange dashed line is the  
prior assumed for the parameter. These two distributions have been 
normalized to the maximum of the a posteriori distribution of the 
eccentricity. Note that the two distributions are very similar, 
indicating that  fitting Doppler-beaming does not add much
constraint to the eccentricity.
Middle panel: as upper panel, but for the argument of periastron 
$\omega$. Lower panel: as upper panel, but for the semi-amplitude of 
the radial velocity modulation $K$.  }
\label{param_distr}
\end{figure}

\begin{figure}[]
\centerline{
\includegraphics[height=9cm,width=8cm,angle=0]{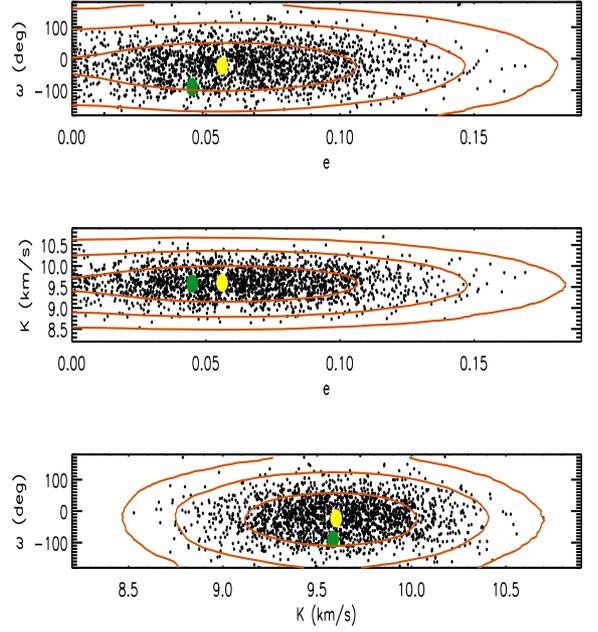}} 
\caption{Upper panel: two-dimensional a posteriori distribution of the argument of periastron $\omega$ vs. the eccentricity $e$ as obtained with the MCMC method. The yellow filled circle corresponds to the best-fit orbital solution of \citet{2011ApJ...730...79J}, while the green circle indicates our best-fit values of the parameters. The orange level lines enclose 68.2, 95, and 99.7 percents of the distribution, respectively. Individual points of the MCMC have been plotted after applying a thinning  factor of 100 to the chain for clarity. Middle panel: as upper panel, but for  the radial velocity semi-amplitude $K$ and the eccentricity $e$. Lower panel: as  upper panel, but for the argument of periastron $\omega$ and the radial velocity semi-amplitude $K$. }
\label{2d-distr}
\end{figure}

Our best fit to the data is plotted in Fig.~\ref{eccentricfit} where the points are  binned for clarity into 40 equal intervals of  mean anomaly $M$. The semi-amplitude of the errorbar of each binned point is the standard error of the flux in the given bin. The posterior distributions of the parameters are plotted in Fig.~\ref{param_distr}. 
The intervals  enclosing between 15.9 and 84.1 percents of the distributions are $ 0.035 \leq e \leq 0.097$; $ -77\fdg14 \leq \omega \leq 33\fdg0$; and $9.290 \leq K \leq 9.897$~km~s$^{-1}$. The correlations among the parameters are not particularly significant, as shown by the two-dimensional posterior distributions plotted in Fig.~\ref{2d-distr}. The best-fit parameters of \cite{2011ApJ...730...79J} fall within the 68.2 percent confidence regions of our two-dimensional distributions. 

The good agreement between the data and the model demonstrates that most of the light modulation of LHS~6343~A can be accounted for by a Doppler-beaming effect with a  fitted semi-amplitude of $1.963 \times 10^{-4}$ in relative flux units. The contribution of the ellipsoidal effect is very small, with a relative semi-amplitude of only $3.05 \times 10^{-6}$ as derived by Eqs.~(\ref{elip1}), (\ref{elip2}), and~(\ref{elip3}). 

\begin{figure}[]
\centerline{
\includegraphics[height=9cm,angle=90]{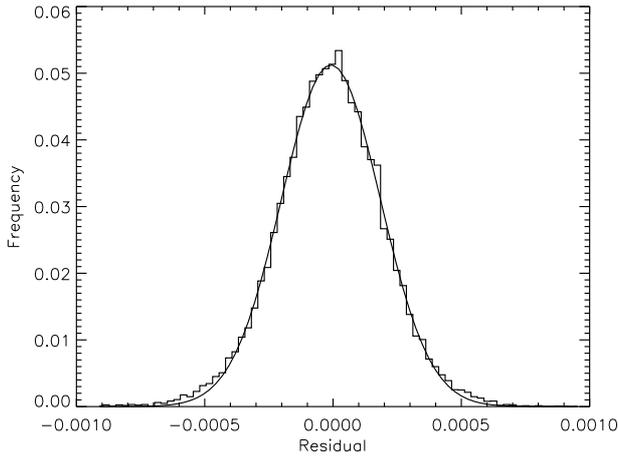}} 
\caption{Distribution of the residuals resulting from the Doppler-beaming plus ellipsoidal effect model of the photometric data of LHS~6343~A
 The solid line is a Gaussian fit to the  distribution.}
\label{hist_resid}
\end{figure}

The distribution of the residuals to our best fit is plotted in Fig.~\ref{hist_resid}. It can be fitted by a Gaussian of standard deviation $ 1.929 \times 10^{-4}$ in relative flux units, although there is an  excess of residuals larger than $\sim 4 \times 10^{-4}$ in absolute value.  The amplitude of the Doppler-beaming plus ellipsoidal modulation is comparable with the standard deviation of the residuals. This accounts for the quite extended confidence intervals found in the parameter distributions. In other words,  the parameters derived by fitting the Doppler-beaming are  of lower accuracy than those derived by fitting the spectroscopic orbit because the radial velocity measurements are more accurate. The a posteriori distributions of the fitted parameters in Fig.~\ref{param_distr} are dominated by their priors, confirming that Doppler-beaming data do not add much information on the model parameters. As a consequence,  the best-fit value of  $\omega$ deviates by more than one standard deviation from the mean of its posterior distribution.

\begin{figure}[]
\centerline{
\includegraphics[height=10cm,angle=90]{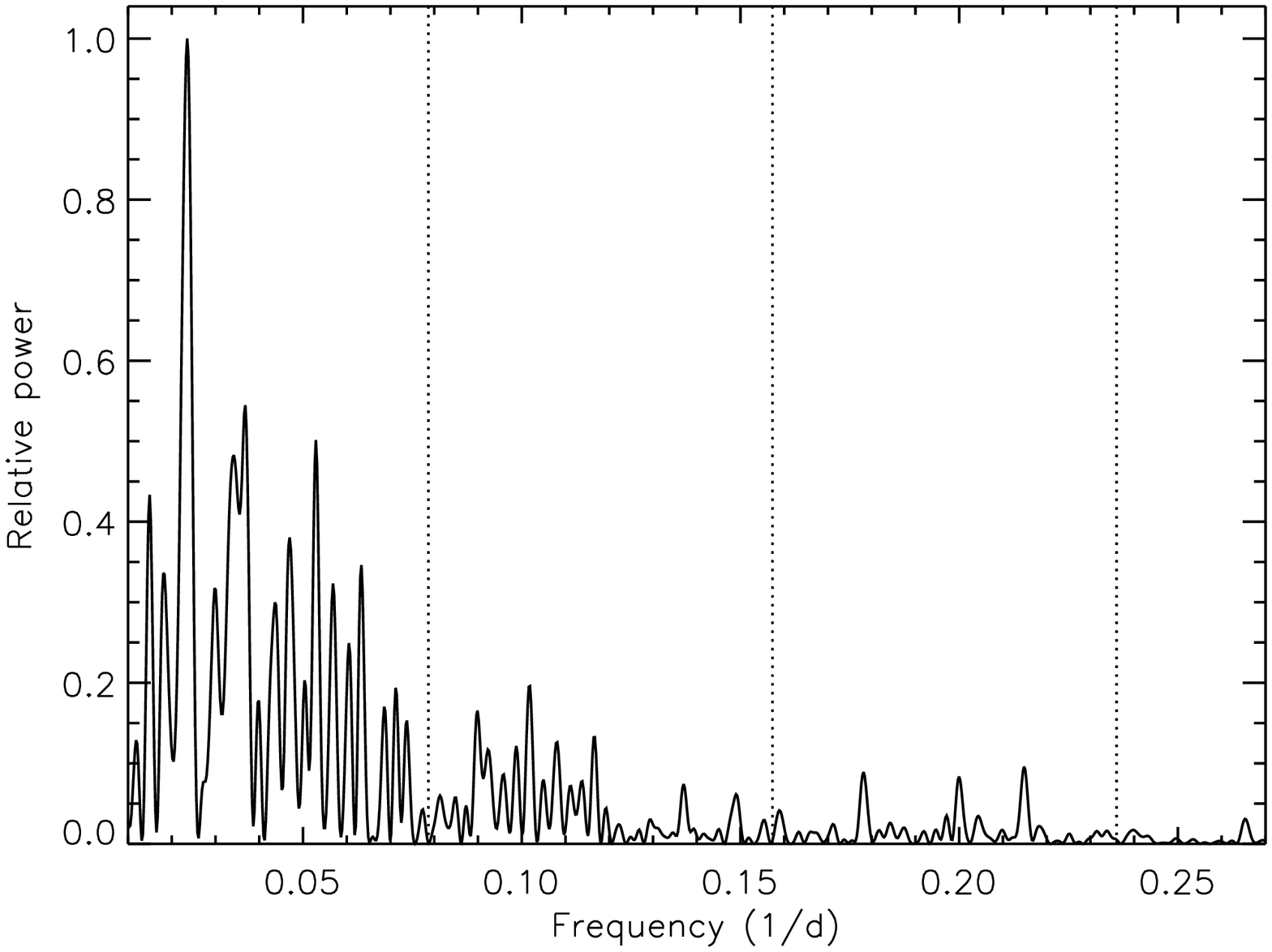}}  
\caption{Lomb-Scargle periodogram of the residuals resulting from the Doppler-beaming plus ellipsoidal effect model of the photometric data of LHS~6343~A. Dotted lines correspond to the frequency of the orbital period and its harmonics.
}
\label{periodogram}
\end{figure}

Finally, we plot in Fig.~\ref{periodogram} the Lomb-Scargle periodogram of the residuals computed with the algorithm of \cite{Press89}. We found a significant periodicity of $42.49 \pm 3.22$ days with a false-alarm probability (FAP) of $4.8 \times 10^{-4}$ as derived by analysing $10^{5}$ random permutations of the flux values with the same time-sampling. The second peak in the periodogram is not a harmonic of the main peak and has an FAP of 20.4 percent, thus it is not considered to be reliable.  The vertical dotted lines indicate the frequencies corresponding to the orbital period and its harmonics. Note that the signal at these frequencies has been almost completely removed by subtracting our model. 


\begin{figure}[]
\centerline{
\includegraphics[height=9.5cm,angle=90]{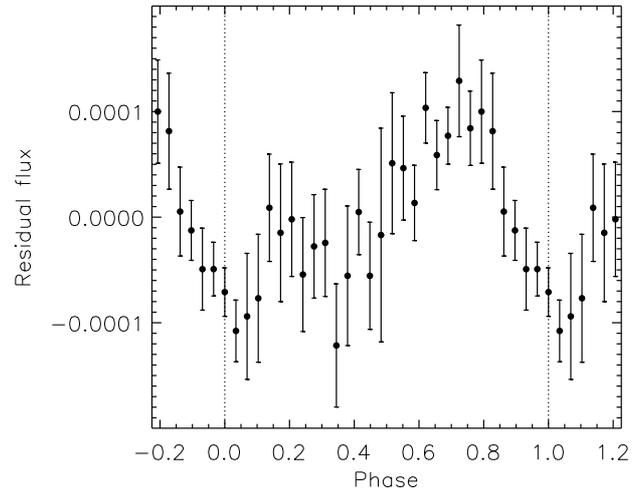}}  
\caption{Residuals of the Doppler-beaming plus ellipsoidal effect model of LHS~6343~A folded at a periodicity of 42.49 days.
}
\label{phasedres}
\end{figure}

The residuals folded at the periodicity of 42.49 days are displayed in Fig.~\ref{phasedres}, showing the mean residual flux vs. phase in 40 equal bins. A  modulation is clearly apparent, suggesting that the primary star's rotation combined with the presence of persistent starspots  might be producing this signal. The possibility that the modulation is due to pulsations seems unlikely given the  long period, but cannot be completely ruled out \citep[see, e.g., ][ and references therein]{Toma72,PallaBaraffe05}. Given the non-sinusoidal shape of the modulation, we have also considered a phasing of the residuals with half the main period (i.e., 21.102 days), but the dispersion of the points around the mean modulation is remarkably higher. 

The non-synchronous rotation of the primary and the eccentricity of the orbit are consistent with  time scales of tidal synchronisation and circularisation at least of the order of the main-sequence lifetime of the system as discussed by \citet{Herrero13}. 

\section{Conclusions}

We have demonstrated that the main assumption made by \cite{Herrero13}  to explain the flux variability in LHS~6343~A as caused by the rotational modulation of photospheric active regions is incomplete. The mean amplitude and phase lag of the modulation can be accounted for by a Doppler-beaming model that  agrees with the  orbital parameters derived by  \cite{2011ApJ...730...79J} by fitting the transits and the radial velocity observations. The ellipsoidal effect was found to be virtually negligible and the reflection effect was not considered given the distance and the luminosity ratio of the two components in the system. 

The periodogram of the residuals reveals a significant periodicity  at $\sim 42.5 \pm 3.2$ days (FAP of $4.8 \times 10^{-4}$),  probably related to the rotation period of  LHS~6343~A.
A more accurate data-detrending  procedure, as is expected to be applied to the final Kepler data release, might be useful to confirm this point and extract more results from the residual analysis.

\begin{acknowledgements} 
The authors are grateful to an anonymous referee for valuable comments that helped to improve their analysis. 
The interpretation presented in this work was originally suggested by one of us (ACC) during a seminar held at the School of Physics and Astronomy of the University of St.~Andrews, Scotland. 
 E. H. and I. R. acknowledge Þnancillary support from
the Spanish Ministry of Economy and Competitiveness (MINECO) and the
''Fondo Europeo de Desarrollo Regional'' (FEDER) through grant AYA2012-
39612-C03-01. C. J. acknowledges the support by the /MINECO/ (Spanish Ministry of 
Economy) - FEDER through grant AYA2009-14648-C02-01, AYA2010-12176-E, 
AYA2012-39551-C02-01 and CONSOLIDER CSD2007-00050. E.~H. is supported by a JAE Pre-Doc grant (CSIC). 
\end{acknowledgements}


\bibliographystyle{aa} 

\end{document}